\documentclass[%
reprint,
nofootinbib,
amsmath,amssymb,
aps,
showkeys,
prd,
]{revtex4-2}

\usepackage{graphicx}% Include figure files
\usepackage{dcolumn}% Align table columns on decimal point
\usepackage{bm}% bold math
\usepackage{mathrsfs}
\usepackage{xcolor}

\def\bIII{{\mathcal{B}^1_2}}
\def\HI{{\mathcal{H}_1}}
\def\HII{{\mathcal{H}_2}}

\def\bhat{{\hat{\mathcal{B}}^1_2}}

\begin{document}

\title{\textit{B-B} plot: A Tool for Accurate Model Selection Using Bayes factors}%

\author{Ankur Barsode}
\email{ankur.barsode@icts.res.in}
\affiliation{International Centre for Theoretical Sciences, Tata Institute of Fundamental Research, Bangalore 560089, India}%

\begin{abstract}
A common task in physics and astronomy is studying which of the competing hypotheses the data prefer. This is usually done by computing the Bayes factor between the two hypotheses, and either interpreting it in terms of the posterior odds or as a ranking statistic for a frequentist p-value test. Here we describe a relationship between the Bayes factor and its distributions under the two competing hypotheses, called the Bayes factor-Bayes factor (\textit{B-B}) relationship, expressed as a diagnostic plot. Using examples from gravitational wave (GW) astronomy, we demonstrate how the \textit{B-B} plot can validate the accuracy of Bayes factor calculations. The \textit{B-B} relationship may also be useful for estimating background distributions of the Bayes factor at low computational cost, even analytically in some cases. We apply this technique in the context of wave-optics lensing of GWs, extrapolating the background distribution from GWTC4 to put a rough bound of $\lesssim 4.1\sigma$ on the statistical significance of the lensing hypothesis of GW231123.
\end{abstract}

%Use showkeys class option if keyword
                      %display desired
\maketitle

%\tableofcontents

\section{Introduction}
\label{sec:introduction}
Model selection is a central task in many areas of physics and astronomy, where one seeks to determine which among the proposed hypotheses are consistent with the observations, and which must be rejected. It lies at the heart of most major scientific discoveries that confirmed or challenged prevailing scientific theories~\cite{abbott2016observation, cms2012observation, abbott2017gw170817, yu2025stochastic, riess1998observational, adame2025desi}. In modern applications, this task is typically statistical in nature and often involves comparing the relative plausibility of models given noisy observations.

In the Bayesian approach, we compute the posterior odds in favor of one hypothesis over another. These are given by the product of the prior odds with the Bayes factor, where the latter is the ratio of evidences of data under each hypothesis. These odds are then compared against an empirical scale (e.g., Jeffrey's~\cite{jeffreys1939theory}) for interpretation.

On the other hand, the frequentist approach seeks to reject a ``null'' hypothesis based on a test ``statistic'' whose value is compared with its ``background'' distribution under the null hypothesis. The statistical significance is then expressed using the false positive probability (FPP), often quoted in terms of Gaussian significance; with $\geq 5\sigma$ usually considered a decisive detection. It is imperative that the statistic be appropriately chosen to maximize the efficiency of detecting the phenomenon of interest, while simultaneously minimizing false alarms. Such a statistic happens to be the Bayes factor itself~\cite{neyman1933ix, fowlie2023neyman}.

Thus, accurate Bayes factors provide the basis for detection under both of these approaches. However, they may be challenging to calculate exactly due to the complexity of realistic models and limited computational resources. Approximations are often made, which raises a question of how correct they really are. Also, as human researchers, we must acknowledge the possibility that our calculations could have errors.

If the (numerically) exact calculation of the Bayes factor using, e.g., nested sampling~\cite{skilling2006nested, sivia2006data, fowlie2020nested}, is possible, only computationally expensive, we may consider validating our approximation of the Bayes factor in a few, typical examples by comparing it against the exact answer. However, if such ground truth is too expensive to obtain, we must resort to internal consistency checks, such as those based on theorems by Alan Turing (the Good checks~\cite{good1965list, sekulovski2024good, okada2026optimality}) or other simulation based checks~\cite{modrak2023simulation, modrak2025simulation}.

\cite{barsode2026lensing} proposed a simulation based validation approach called the Bayes factor-Bayes factor or \textit{B-B} plot, and used it to test a Bayesian method to search for strong gravitational lensing of gravitational waves (GWs). They also used it to extrapolate background distributions of the GW strong lensing Bayes factors to forecast detection efficiency in upcoming observing scenarios.

In this paper, we discuss the applications of the \textit{B-B} plot in more detail. We provide a procedure to generate a \textit{B-B} plot for a given Bayesian pipeline, discuss the conditions under which it should be considered valid, and mention related caveats. We also define a scalar ``badness'' metric to assess whether the \textit{B-B} plot is satisfied by a given Bayesian pipeline.

Using a pedagogical example of gravitational waveform distortion searches, we show how the \textit{B-B} plot can provide a direct and internally consistent diagnostic for assessing the accuracy of approximate Bayes factor calculations, without requiring access to ground-truth results. The \textit{B-B} plot can also guide the systematic improvement of such approximations. As a concrete example of the latter, we improve upon an existing GW strong lensing search pipeline, making it more accurate as well as more computationally efficient.

We also solidify and generalize the background estimation technique used by \cite{barsode2026lensing}, showing how the \textit{B-B} plot can help save computational costs of frequentist FPP estimation. We also present semi-analytical background extrapolations to regimes inaccessible to brute-force simulations, and apply them to place bounds on the statistical significance of the lensing hypothesis of GW231123~\cite{abac2025gw231123}.

This paper is structured as follows: in Sec.~\ref{sec:bb_plot} we provide a primer on Bayesian analysis and the \textit{B-B} plot. In Sec.~\ref{sec:benchmarking}, we demonstrate benchmarking using \textit{B-B} plot in two model selection problems in GW astronomy: waveform distortion and strong gravitational lensing. In Sec.~\ref{sec:background_estimation}, we estimate backgrounds for gravitational lensing of GWs using the \textit{B-B} relationship and estimate an upper bound on the statistical significance of the lensing hypothesis of GW231123. Our conclusions are presented in Sec.~\ref{sec:conclusion}.

\section{The Bayes factor-Bayes factor (\textit{B-B}) Plot}
\label{sec:bb_plot}

At the heart of Bayesian model selection lies the Bayes theorem
\begin{equation}
\label{eq:general_Bayes_theorem}
P(\vec{\theta}_i \mid d, \mathcal{H}_i) = \dfrac{P(d \mid \vec{\theta}_i, \mathcal{H}_i) ~ P(\vec{\theta}_i \mid \mathcal{H}_i)}{P(d \mid \mathcal{H}_i)}
\end{equation}
which, for a model hypothesis $\mathcal{H}_i$, expresses the posterior distribution of model parameters $\vec{\theta}_i$, given data $d$, in terms of the likelihood of getting the data given those parameters $P(d \mid \vec{\theta}_i, \mathcal{H}_i)$, and their prior distribution $P(\vec{\theta}_i \mid \mathcal{H}_i)$. The denominator on the right is the Bayesian evidence that normalizes the posterior
\begin{equation}
\label{eq:general_evidence}
P(d \mid \mathcal{H}_i) = \int \mathrm{d}\vec{\theta}_i ~ P(d \mid \vec{\theta}_i, \mathcal{H}_i) ~ P(\vec{\theta}_i \mid \mathcal{H}_i)
\end{equation}

The Bayes factor is defined as the ratio of evidences under different model hypotheses, say, for $i=1,2$,
\begin{equation}
\label{eq:general_Bayes_factor}
\bIII(d) = \dfrac{P(d \mid \HI)}{P(d \mid \HII)}.
\end{equation}
It is often called the \textit{likelihood ratio} of the competing models for producing the observed data. Note that the Bayes factor also implicitly depends on the prior choices under each hypothesis, apart from being an explicit function of the data.

It is useful to clarify what is meant by ``data'' in this context. The raw numbers recorded by the instruments typically pass through a pipeline of processes that reduce, refine, and simplify them. For a given observation, the data product at the end of every intermediate process in this pipeline should, in principle, give the same Bayes factor. Ultimately, we might consider the calculation of the Bayes factor itself as a process that returns ``data'' (= the Bayes factor) using which we could compute the Bayes factor. Thus
\begin{equation}
\label{eq:general_BB}
\bIII = \dfrac{P(\bIII \mid \HI)}{P(\bIII \mid \HII)},
\end{equation}
where the Bayes factor is being treated as a random variable, and the equality holds at the level of its probability densities under each hypothesis.

More formally, following \cite{barsode2026lensing}, we can derive the above relationship by noting that the probability density of the Bayes factor is given by marginalizing over all possible data that result in that Bayes factor. For the hypothesis $\HI$, this is
\begin{equation}
\label{eq:likelihood_transform}
P(\bIII \mid \HI) = \int \mathrm{d}d ~ P(d \mid \HI)~\delta\left(\bIII-b(d)\right)
\end{equation}
where $\delta$ is the Dirac delta function, and we are temporarily denoting the Bayes factor function by $b(d)$ to avoid confusion with the Bayes factor as a random variable $\bIII$. Using the definition in Eq.~\eqref{eq:general_Bayes_factor}, we can substitute $P(d \mid \HI)$ by $b(d) \cdot P(d \mid \HII)$
\begin{equation}
P(\bIII \mid \HI) = \int \mathrm{d}d ~ b(d) ~ P(d \mid \HII)~\delta\left(\bIII-b(d)\right).
\end{equation}
The Dirac delta function restricts the integral to those data realizations for which $b(d)=\bIII$. On this support, $b(d)$ can therefore be treated as a constant within the integral. This gives
\begin{equation}
P(\bIII \mid \HI) = \bIII ~ \int \mathrm{d}d ~ P(d \mid \HII)~\delta\left(\bIII-b(d)\right).
\end{equation}
The integral is analogous to Eq.~\eqref{eq:likelihood_transform}, only in this case it gives the distribution of Bayes factors under the hypothesis $\HII$, resulting in
\begin{equation}
\label{eq:general_BB_TuringGood}
P(\bIII \mid \HI) = \bIII ~ P(\bIII \mid \HII),
\end{equation}
which is identical to Eq.~\eqref{eq:general_BB}. Thus, the ratio of the PDFs of $\bIII$ under the competing hypotheses at various values of $\bIII$ should be equal to the corresponding $\bIII$.

We will refer to this relationship between the Bayes factor and its distributions under the competing hypotheses as the Bayes factor-Bayes factor or \textit{B-B} relationship. Multiplying both sides of Eq.~\eqref{eq:general_BB_TuringGood} by the $n$th power of $\bIII$ and integrating over $\bIII$, we recover the Turing-Good theorem that the expectation value of the $n$th power of $\bIII$ under hypothesis $\HI$ is equal to that of the $(n+1)$th power of $\bIII$ under $\HII$. In particular, $n=0$ gives the rather surprising result that the average value of the Bayes factor under the ``null'' (i.e., $\HII$) hypothesis is 1. Previous work proposed to use these expectation values for diagnostics~\cite{sekulovski2024good, okada2026optimality}, though there has been some criticism~\cite{modrak2025simulation}.

The \textit{B-B} relationship can directly be used as a diagnostic test by making a \textit{B-B} plot: a graph of the right-hand side of Eq.~\eqref{eq:general_BB} against its left-hand side. It is expected to lie on the diagonal equality line. There may be fluctuations due to finite sampling noise, but these can be estimated using, e.g., bootstrapping methods.

Practically, we perform two separate experiments: first, we generate random realizations of data from the prior corresponding to the first hypothesis. For each such realization, we compute the Bayes factor, effectively obtaining a large number of samples from which the distribution $P(\bIII \mid \HI)$ can be empirically reconstructed using, e.g., histograms. In the second experiment, we repeat the same procedure but with data generated using the prior corresponding to the second hypothesis to empirically construct $P(\bIII \mid \HII)$. Since the data in each experiment is generated from its respective prior, and \textit{both} the priors enter the calculation of Bayes factor, it should not be too surprising that the distributions $P(\bIII \mid \HI)$ and $P(\bIII \mid \HII)$, though obtained using data from two independent experiments, are nevertheless related to each other. Equation~\eqref{eq:general_BB} is precisely that relationship, and is visualized by plotting the ratio of these empirical reconstructions against $\bIII$.

We can also construct a scalar diagnostic to measure the badness of a \textit{B-B} plot, accounting for finite sample variance. The empirical mean vector $\mu$ and covariance matrix $\Sigma$ of the residuals $\log_{10} \left[(P(\bIII \mid \HI) / P(\bIII \mid \HII)) / \bIII \right]$ can be estimated using techniques such as bootstrapping, or, for histogram-based density reconstructions, using multinomial distributions. In the latter case, with $k$ histogram bins, we assume that both $k$ and the bin counts are large enough that the residuals can be assumed to follow a k-dimensional Gaussian distribution\footnote{though we have not verified this explicitly for the finitely sampled distributions tested here}. Under this multivariate Gaussian assumption, the quantity $\mu^T\Sigma^{-1}\mu$ would be distributed as a $\chi^2$ with mean $k$. A natural measure of the badness of the \textit{B-B} plot is given by the total probability of obtaining residuals further away from 0 than what we have observed, or, equivalently, by the reversed cumulative distribution function (CDF) of the $\chi^2$ at $\mu^T\Sigma^{-1}\mu$. Throughout this paper, we use the following score as our badness metric
\begin{equation}
\label{eq:badness}
b = \sqrt{2} ~ \mathtt{erfc}^{-1} \left[ 1 - C_{\chi^2}(\mu^T\Sigma^{-1}\mu, k) \right].
\end{equation}
Here, we have used the inverse of the complementary error function to convert the reversed CDF into Gaussian significance levels for easy interpretation. Smaller values ($\lesssim 3\sigma$) of $b$ indicate good \textit{B-B} plots.

Bayesian model selection involves interpreting the Bayes factor at face value, necessitating accuracy. On the other hand, frequentist model selection requires expensive simulations and data processing to generate the ``background'' $P(\bIII \mid \HII)$ and ``foreground'' $P(\bIII \mid \HI)$ distributions for estimating the False Positive (FPP) and False Dismissal (FDP) probabilities, respectively. The \textit{B-B} plot connects the quantities these two approaches rely upon, making it possible to use one for obtaining the other. In subsequent sections, we will show how the \textit{B-B} plot can be used to benchmark the accuracy of Bayes factors using simulated backgrounds and foregrounds. If a validated Bayesian methodology exists, we show how the \textit{B-B} plot can help estimate its background at low computational cost.

\section{Benchmarking Bayes Factors}
\label{sec:benchmarking}
The Bayes factor is analytically tractable in only a few, limited cases; realistic data and models are often too complicated to enable a direct evaluation of the evidences in Eq.~\eqref{eq:general_evidence}. While recent developments in nested sampling and computing hardware have made it possible to evaluate evidences in several real-life problems, it still remains relatively expensive, and we often resort to making approximations to simplify the calculations at the risk of incurring some error.

The \textit{B-B} plot can help us quantify such errors. If it is possible to simulate data under each of the competing hypotheses, we could generate a large number of such data, evaluate our approximation $\bhat$ of the Bayes factor on them, and empirically construct the distributions $P(\bhat \mid \HI)$ and $P(\bhat \mid \HII)$. If our approximation is correct, a \textit{B-B} plot of $P(\bhat \mid \HI) / P(\bhat \mid \HII)$ against $\bhat$ will lie close to the equality line, while deviations may help identify potential discrepancies.

We caution, however, that satisfying the \textit{B-B} plot alone is not sufficient to deem the Bayes factor calculation as correct. A trivial counterexample is when we define $\bhat =  P(S \mid \HI) / P(S \mid \HII)$, where $S$ can be an arbitrary scalar function of the random realizations of data. Multiplying and dividing this equation by $dS/d\bhat$, it is easy to recognize that this will satisfy Eq.~\eqref{eq:general_BB}, even though $\bhat$ may be completely different from the true Bayes factor one would obtain from Eq.~\eqref{eq:general_Bayes_factor}.

Nevertheless, while the efficiency may be degraded relative to that corresponding to the true Bayes factor, $\bhat$ is still interpretable in the Bayesian sense of relative odds. Also, if we are systematically approximating the exact expression of the Bayes factor (rather than working with the likelihood ratio of an arbitrary statistic), it appears (though we provide no rigorous proof) that such a caveat may be unimportant.

Another important point to note is that making \textit{B-B} plots requires the data to be drawn from the same prior distributions as those used in the computation of the Bayes factors. Any mismatches should be corrected, e.g. using reweighting. If selection effects are present, they should be applied consistently to the data and the prior. If these conditions are not met, the \textit{B-B} may not meet the goodness criteria defined above even though the Bayes factor calculation may be correct (see Appendix~\ref{sec:dingo_lensing} for an example). In all the applications discussed below, we ensure that the data are generated from the same (selection effects included) priors as those used in the Bayes factor computation.

\subsection{Gravitational Waveform Distortion}
\label{sec:benchmarking_AGGR}

\begin{figure*}
\centering
\includegraphics[width=\linewidth]{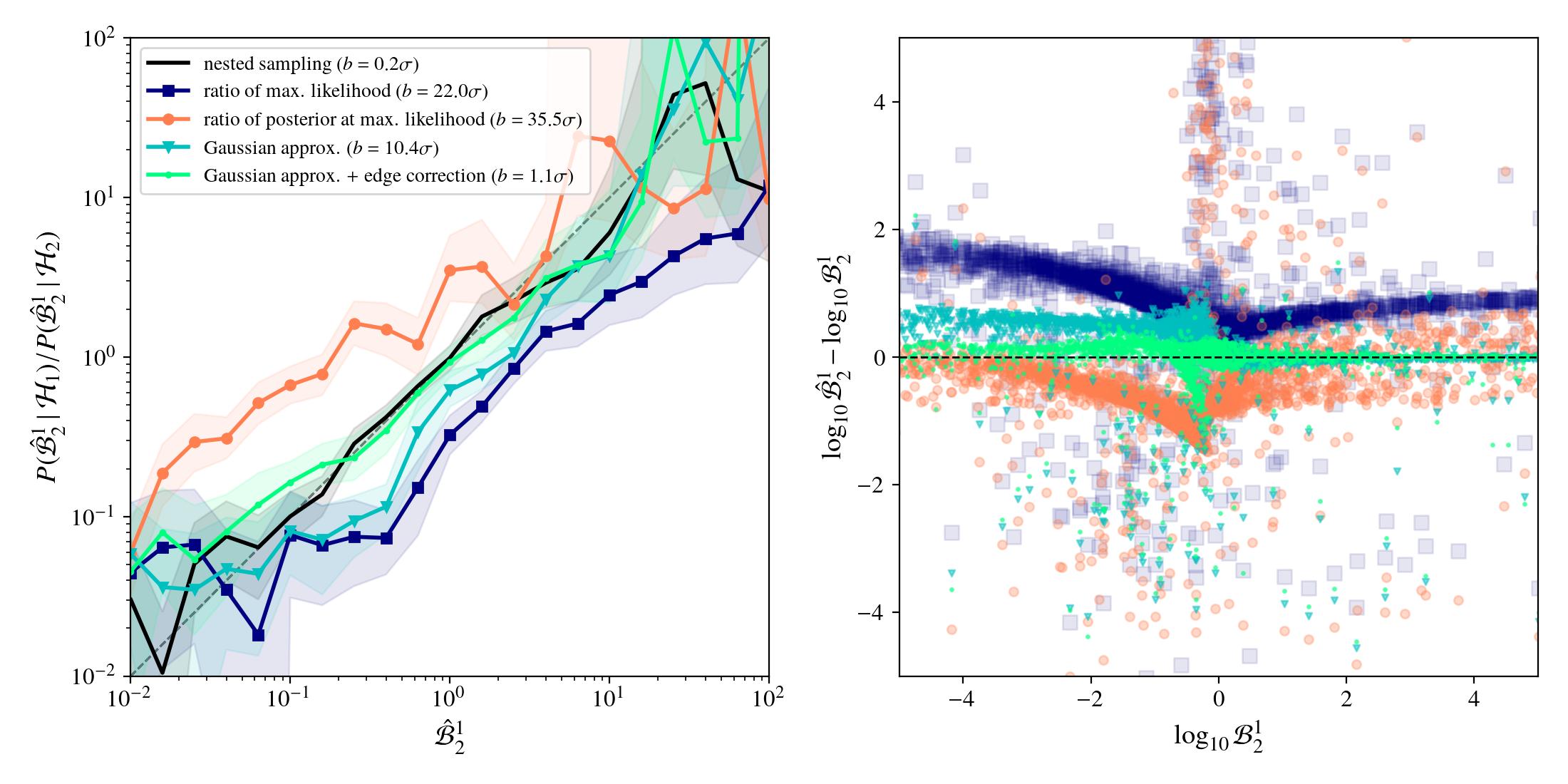}
\caption{Left: the \textit{B-B} plots for (numerically) exact nested sampling based Bayes factor between the alternative to GR and GR hypotheses in our toy problem, along with those corresponding to various approximations described in Eqs.~\eqref{eq:sgbf_approx_ML}, \eqref{eq:sgbf_approx_MLP}, \eqref{eq:sgbf_approx_G}, and \eqref{eq:sgbf_approx_GE}. Their badness (see Eq.~\eqref{eq:badness}) is also indicated. Shaded regions indicate 90\% credible regions estimated assuming multinomial histogram errors. Right: difference between approximate and exact Bayes factor calculations plotted against the latter, for the four estimates discussed in the left-hand panel.}
\label{fig:sgbf_approximations}
\end{figure*}

One of the motivations behind GW astronomy is to test the theory of General Relativity (GR) in the extreme conditions around a compact binary merger. Any deviations would end up distorting the waveforms of the GWs from our expectations under GR (see, e.g., \cite{cornish2011gravitational}). Such distortions could also be caused by gravitational lensing of GWs \cite{lawrence1971focusing, takahashi2003wave, dai2017waveforms}, and we will return to this in Sec.~\ref{sec:GW231123}. In this section, we will focus on a toy model that captures the main spirit of the search for generic waveform distortion effects, and show how the \textit{B-B} plot may assist in improving these searches.

Our problem consists of a fiducial ``GR'' signal
\begin{equation}
\label{eq:sgbf_GR_waveform}
h_\mathrm{GR}(t; A, \sigma) = A~\sin(100t)~e^{-t^2/2\sigma^2}
\end{equation}
given by a Gaussian with amplitude $A$ and width $\sigma$ modulating a sine wave, and an alternative to GR ``AG'' signal
\begin{equation}
\label{eq:sgbf_AG_waveform}
h_\mathrm{AG}(t; A, \sigma, \tau) = \begin{cases}
A\sin(100t)e^{-t^2/2\sigma^2},& \text{if } t\leq 0\\
A\sin(100t)e^{-t^2/2\sigma^2}~e^{-t/\tau},& \text{otherwise}
\end{cases}
\end{equation}
having an additional exponential decay with timescale $\tau$ beyond the peak amplitude, standing in for the waveform distortion. For $\tau \gg \sigma$, the distortion is small, and the AG and GR signals are almost identical.

The priors describing these signals' parameters are assumed to be loguniform with ranges $A\in[0.01,10]$, $\sigma\in[0.1,10]$ for both GR and AG models, with $\tau\in[0.1,10]$ in the AG model. Only those regions of the prior space that result in optimal signal-to-noise ratio (SNR) $\rho_\mathrm{opt} \geq 8$ in standard normal white noise are considered, where $\rho_\mathrm{opt}$ is defined by
\begin{equation}
\label{eq:sgbf_optimal_SNR}
\rho_\mathrm{opt} = \sqrt{\int_{-\infty}^{\infty} \mathrm{d}t~h_i^2(t; \vec{\theta}_i)}
\end{equation}
where $h_i$ and $\vec{\theta}_i$ stand for the waveform and parameters of a particular model, $h_1=h_\mathrm{AG}$ and $\vec{\theta}_1=\{A,\sigma,\tau\}$ for AG, and $h_2=h_\mathrm{GR}$ and $\vec{\theta}_2=\{A,\sigma\}$ for GR.

For either of these models, we can draw a random parameter point from the appropriate prior, generate a signal according to the waveform models in Eqs.~\eqref{eq:sgbf_GR_waveform} or \eqref{eq:sgbf_AG_waveform}, and inject (i.e., add) it in standard normal white noise to generate a mock data strain $d(t)$. Given $d(t)$, the functional forms of the two waveform models, and the two priors, we would like to identify whether the injected signal was AG or GR.

It is straightforward to set up the Bayesian model selection formalism. The log-likelihood is given by
\begin{equation}
\label{eq:sgbf_log_likelihood}
\ln P(d \mid \vec{\theta}_i) = -\ln{2\pi} -\frac{1}{2} \int_{-\infty}^{\infty} \mathrm{d}t~ (d(t) - h_i(t; \vec{\theta}_i))^2.
\end{equation}
with $i=1$ for the AG model and $2$ for the GR model, and we have dropped the explicit dependence on $\mathcal{H}_i$ since $\vec{\theta}_i$ already fixes it. This, combined with the priors defined above, can be fed into a nested sampler such as, e.g., \textsc{Nautilus}~\cite{nautilus}, to sample the posterior distributions (Eq.~\eqref{eq:general_Bayes_theorem}) and calculate the evidences (Eq.~\eqref{eq:general_evidence}) under each of the two hypotheses, which would then give the Bayes factor
\begin{equation}
\label{eq:sgbf_Bayes_factor}
\bIII = \dfrac{\int \mathrm{d}\vec{\theta}_1~P(d \mid \vec{\theta}_1) P(\vec{\theta}_1 \mid \HI)}{\int \mathrm{d}\vec{\theta}_2~P(d \mid \vec{\theta}_2) P(\vec{\theta}_2 \mid \HII)}.
\end{equation}

However, while the problem posed above is simple enough, most realistic gravitational waveform distortion effects require sampling 15 or higher dimensional likelihoods, involving millions of likelihood evaluations that take hundreds to thousands of CPU hours. It is worth exploring whether simple approximations can be made to reduce these costs, and whether those approximations can be validated without ever computing the full nested sampling based $\bIII$.

Therefore, in the beginning, we will assume that even for the simple problem posed above, it is impossible to sample the posteriors, compute the evidences, and calculate the $\bIII$ in a brute force manner. We will assume, however, that it is still possible to find the parameter vectors $\vec{\theta}_1^\mathrm{ML}$ and $\vec{\theta}_2^\mathrm{ML}$ corresponding to the maximum likelihood of the two models, and the Fisher covariance matrices~\cite{cutler1994gravitational} $\Sigma_1$ and $\Sigma_2$ around those locations. At low SNRs, this may be challenging as the posterior may be significantly non-Gaussian and possibly multimodal in the usual physical parameter space. However, clever reparameterization (such as that proposed in~\cite{roulet2022removing}) may simplify the posterior shape enough to obtain these quantities with relatively few likelihood evaluations.

The program is as follows: we simulate 3000 random realizations of data from each of the AG and GR models. On each sample strain, we maximize the likelihood within the prior region using \texttt{scipy.optimize.minimize}, starting from an initial guess $(A,\sigma,\tau) = (1,1,1)$ for AG and $(A,\sigma) = (1,1)$ for GR. Around the respective maximum likelihood points, we compute the Fisher covariance matrices using finite-difference derivatives. Using these quantities, we evaluate our approximation $\bhat$ of the Bayes factor (see below). With these 3000 $\bhat$ samples, we empirically reconstruct $P(\bhat \mid \HI)$ and $P(\bhat \mid \HII)$ and make a \textit{B-B} plot. If it lies close to equality, we deem the approximation valid; else, we improve upon it with additional terms.

We consider four approximations for the Bayes factor
\begin{enumerate}
\item Ratio of maximum likelihoods
\begin{equation}
\label{eq:sgbf_approx_ML}
\bhat = \dfrac{P(d\mid \vec{\theta}_1^\mathrm{ML})}{P(d\mid \vec{\theta}_2^\mathrm{ML})}.
\end{equation}
\item Ratio of posteriors at maximum likelihood points
\begin{equation}
\label{eq:sgbf_approx_MLP}
\bhat = \dfrac{P(d\mid \vec{\theta}_1^\mathrm{ML})}{P(d\mid \vec{\theta}_2^\mathrm{ML})} ~ \dfrac{P(\vec{\theta}_1^\mathrm{ML} \mid \HI)}{P(\vec{\theta}_2^\mathrm{ML} \mid \HII)}.
\end{equation}
\item Gaussian approximation (also known as Laplace's approximation)
\begin{equation}
\label{eq:sgbf_approx_G}
\bhat = \sqrt{2\pi} \dfrac{P(d\mid \vec{\theta}_1^\mathrm{ML})}{P(d\mid \vec{\theta}_2^\mathrm{ML})} ~ \dfrac{P(\vec{\theta}_1^\mathrm{ML} \mid \HI)}{P(\vec{\theta}_2^\mathrm{ML} \mid \HII)} ~ \dfrac{\lvert \Sigma_1 \rvert^{1/2}}{\lvert \Sigma_2 \rvert^{1/2}}.
\end{equation}
\item Gaussian approximation with edge corrections
\begin{multline}
\label{eq:sgbf_approx_GE}
\bhat = \sqrt{2\pi} \dfrac{P(d\mid \vec{\theta}_1^\mathrm{ML})}{P(d\mid \vec{\theta}_2^\mathrm{ML})} ~ \dfrac{\lvert \Sigma_1 \rvert^{1/2}}{\lvert \Sigma_2 \rvert^{1/2}}\\
\times \dfrac{\left<P(\vec{\theta}_{1,k} \mid \HI)\right>_{\vec{\theta}_{1,k}\sim G_1}}{\left<P(\vec{\theta}_{2,k} \mid \HII)\right>_{\vec{\theta}_{2,k}\sim G_2}} ~ \dfrac{f_1}{f_2}.
\end{multline}
where, for $i\in\{1,2\}$, $G_i$ are multivariate Gaussians with mean $\vec{\theta}_i^\mathrm{ML}$ and covariance $\Sigma_i$, $\vec{\theta}_{i,k}$ are samples from these Gaussians, and $f_i$ is the fraction of those samples that lie within the prior boundaries.
\end{enumerate}
The first two do not resemble expressions of the exact Bayes factor at all, and are chosen to illustrate how such bad approximations may be identified. The latter two are based on the Gaussian approximation and should therefore be treated with more importance.

The \textit{B-B} plots for these approximations are shown in the left-hand panel of Fig.~\ref{fig:sgbf_approximations}. By checking whether the ratio $P(\bhat \mid \HI) / P(\bhat \mid \HII)$ lies above or below the equality line, we can infer that the likelihood ratio is an overestimate, while the posterior ratio is an underestimate. This is expected since they fail to account for the three main constituents of the Bayesian evidence: the likelihood, times the posterior, and their integration over the appropriate region of parameter space. The Gaussian approximation takes these into account and therefore gives a less biased \textit{B-B} plot. Accounting for edge effects further improves the Gaussian approximation, substantially reducing the remaining bias in the \textit{B-B} plot. It may therefore be a viable alternative to expensive nested sampling-based Bayes factor calculation.

While future work can apply the above approximation in more realistic studies of gravitational waveform distortion, the key takeaway of the above exercise is that we arrived at this result \textit{without} having to compare it against nested sampling. Equation~\eqref{eq:general_BB} acts as a consistency test by itself, and as long as we can simulate data from the priors, we can use \textit{B-B} plots as a diagnostic tool to assess the accuracy of Bayes factor calculations without requiring access to the ground truth.

For completeness, we also compute the (numerically) exact Bayes factor using the \textsc{Nautilus}~\cite{nautilus} nested sampler using 1024 live points, run until the fractional evidence in the live set is $\leq 1\%$, or the effective number of samples is $\geq 10^4$. The left-hand panel of Fig.~\ref{fig:sgbf_approximations} shows that its \textit{B-B} plot lies over the diagonal equality line, as expected. Looking at the residuals between it and the approximations discussed above (right hand panel), we find that though the \textit{B-B} plot fails to capture the discrepancies near small $\lvert \log_{10}\bhat \rvert$, their nature (i.e., over- or underestimation) matches what we had inferred above in the regime of interest for conclusive model selection ($\lvert \log_{10}\bhat \rvert \gtrsim 1$).

\subsection{Strong Gravitational Lensing of Gravitational Waves}
\label{sec:benchmarking_PO2}

\begin{figure*}
\centering
\includegraphics[width=0.32\linewidth]{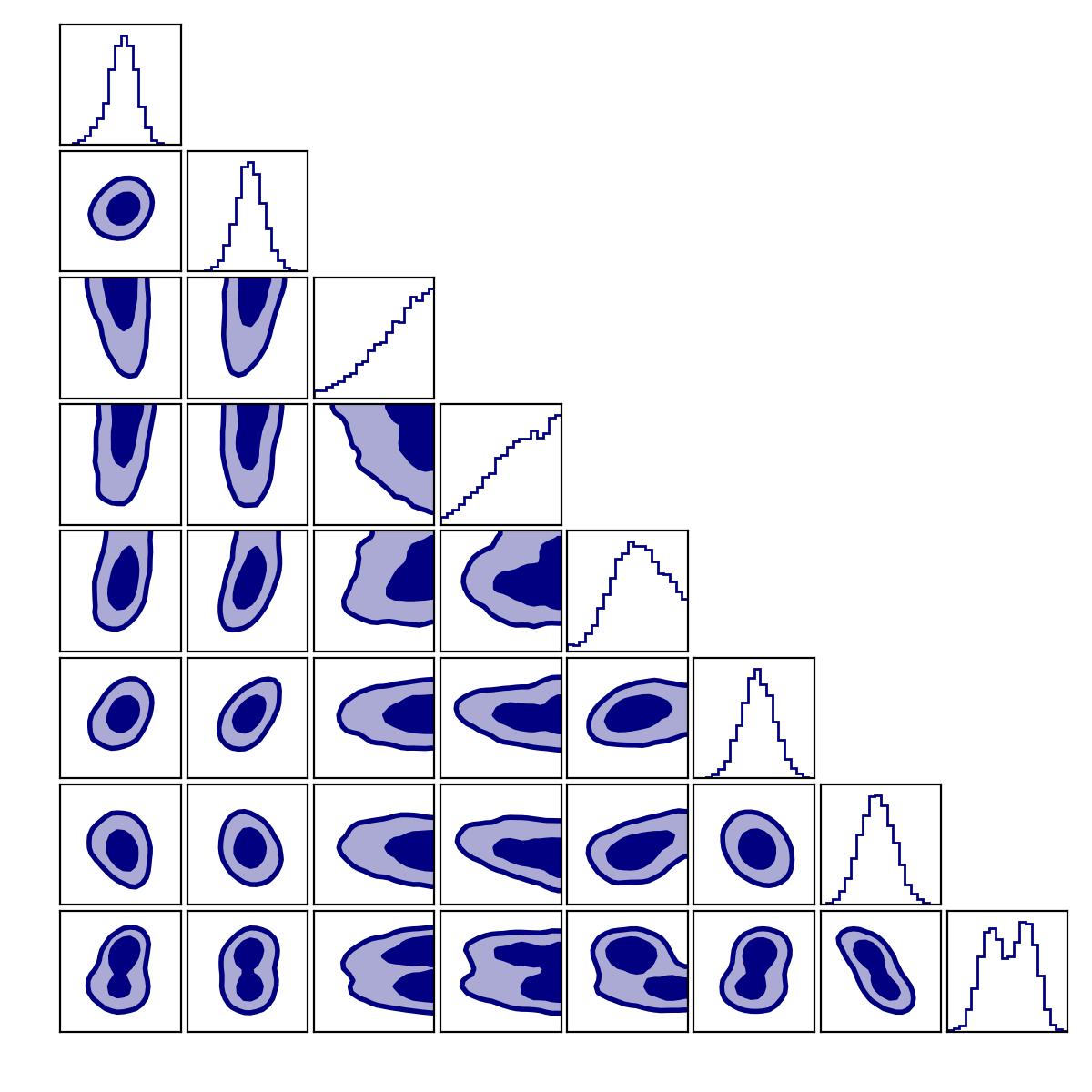}
\includegraphics[width=0.66\linewidth]{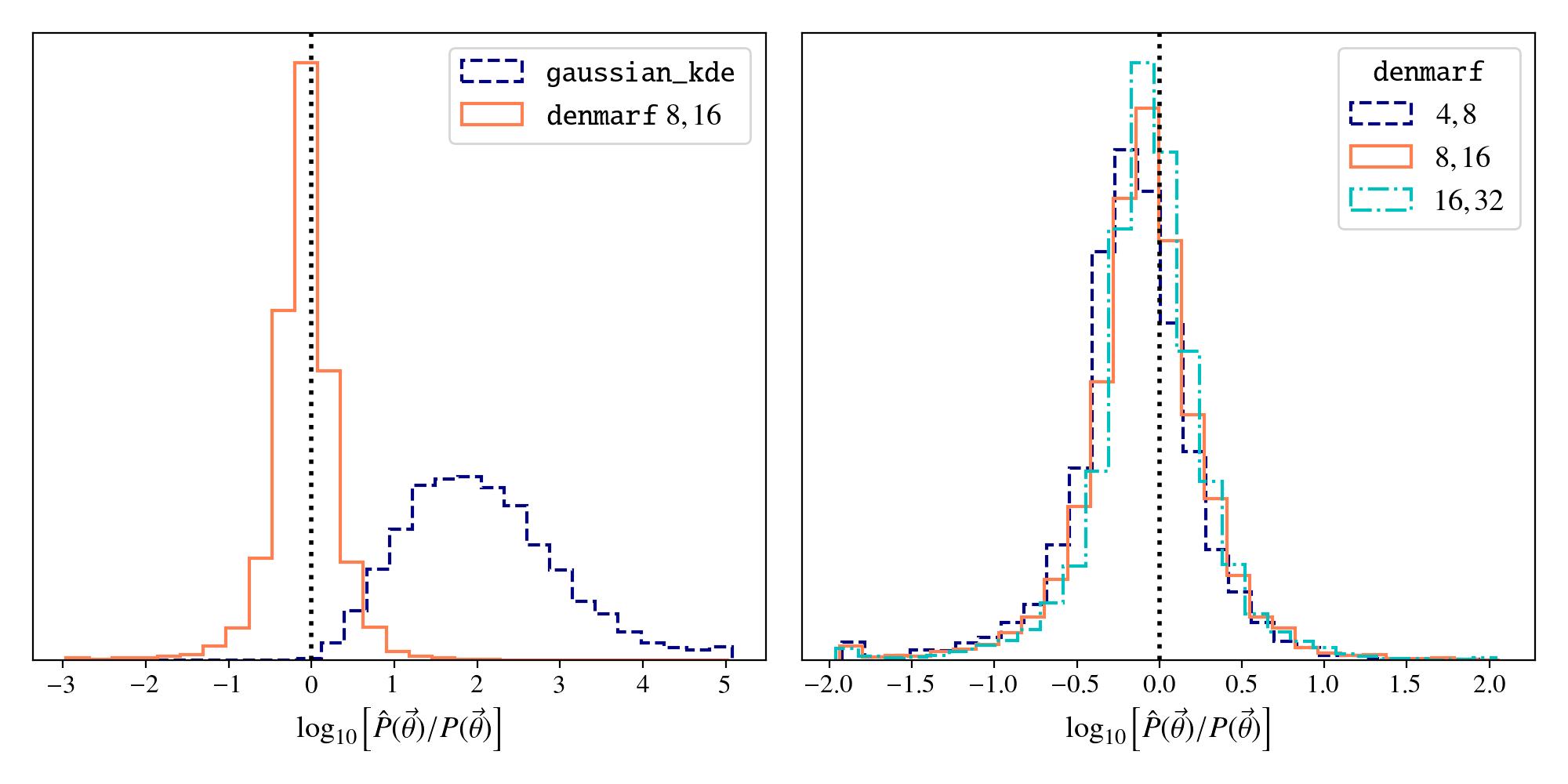}
\caption{The leftmost cornerplot shows an example joint distribution on which we test density estimation methods. This is generated from a random realization of a mixture of 20 multivariate Gaussians, and simulates the qualitative features (such as multivariate, non-linear correlations, multimodality, prior boundary effects, etc.) of the distributions we generally need to reconstruct when calculating GW strong lensing Bayes factors using PO2.0. In the middle and rightmost panels, we show the error in the estimated density $\hat{P}$ relative to the true density $P$ for a set of sample points $\vec{\theta}$. The middle panel compares \texttt{gaussian\_kde} based density estimation against \texttt{denmarf}, while the rightmost panel compares various configurations (number of hidden units, number of blocks) of \texttt{denmarf} against each other.}
\label{fig:test_kde_denmarf}
\end{figure*}

\begin{figure*}
\centering
\includegraphics[width=0.8\linewidth]{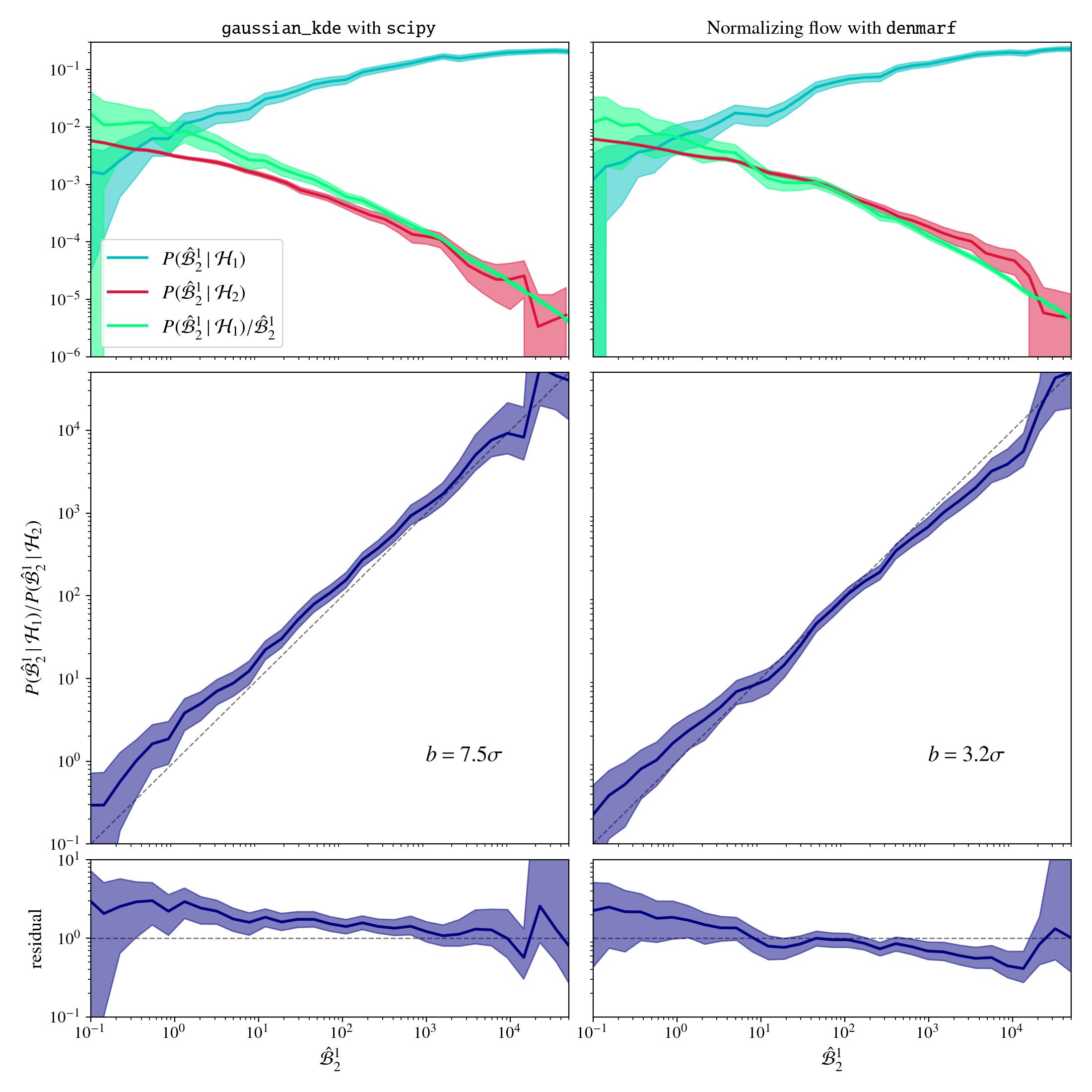}
\caption{The top row shows the probability distribution of the GW strong lensing Bayes factor $\bIII$ obtained using the PO2.0 method from a large number of simulated lensed and unlensed signals. Also plotted is $P(\bIII \mid \HI)$ scaled by $\bIII$ to show that it is very close to $P(\bIII \mid \HII)$. The same information is shown in the middle row as \textit{B-B} plots, with badness (see Eq.~\eqref{eq:badness}) as indicated, while the lower row shows the residual $(P(\bhat \mid \HI) / P(\bhat \mid \HII))/\bhat$. Shaded regions indicate 90\% credible regions estimated assuming multinomial histogram errors. The left column shows results with the older \texttt{gaussian\_kde} based approach, while the right column shows the same using the normalizing flow based implementation of PO2.0 $\bIII$.}
\label{fig:PO2_kde_vs_denmarf_BB_plot}
\end{figure*}

In our second demonstration of the use of \textit{B-B} plots as a diagnostic tool, we describe our success at improving a Bayesian search pipeline for strongly lensed GWs.

Strong gravitational lensing of GWs occurs when a massive object (e.g., a galaxy) lies along our line-of-sight to a GW source (e.g., a binary black hole (BBH)). The gravity of the massive object (i.e., the ``lens'') bends the paths of the GWs, potentially resulting in multiple-- nearly identical-- copies of the same source waveform reaching our detectors at different times. Except in very rare cases, the only way to detect them is by confidently identifying pairs of signals that resemble each other to such a degree that random chance alone cannot explain it~\cite{haris2018identifying}.

The Bayes factor between the lensed and unlensed hypotheses is schematically given by~\cite{harris2020array, janquart2021fast, lo2023bayesian, cheung2023mitigating, janquart2023return, barsode2025fast, hannuksela2025strong}
\begin{equation}
\label{eq:GWSL_BLU_joint_likelihood}
\bIII = \dfrac{\int \mathrm{d}\vec{\theta}_1 ~ P(d_A,d_B \mid \vec{\theta}_1) ~ P(\vec{\theta}_1 \mid \HI)}{\prod\limits_{j\in\{A,B\}}\int \mathrm{d}\vec{\theta}_2 ~ P(d_j \mid \vec{\theta}_2) ~ P(\vec{\theta}_2 \mid \HII)}
\end{equation}
where $d_A,d_B$ are the data corresponding to the two individual signals, $\vec{\theta}_1$ and $\vec{\theta}_2$ parameterize the lensed and unlensed hypotheses $\HI$ and $\HII$ respectively. $\vec{\theta}_2$ are the usual BBH waveform parameters, while $\vec{\theta}_1$ additionally include lensing-related parameters such as magnification, phase shift, and time delay.

The Posterior Overlap 2.0 (hereafter, PO2.0)~\cite{barsode2025fast} formalism evaluates the above using the following approximation
\begin{equation}
\label{eq:GWSL_BLU_joint_likelihood_PO2}
P(d_A,d_B \mid \vec{\theta}_1) \propto P(\vec{\theta}_2 \mid d_A) \cdot P(\vec{\theta}_2+\Delta\theta \mid d_B)
\end{equation}
i.e., the joint likelihood is proportional (up to priors) to the product of the individual unlensed posteriors, with the second signal's posterior evaluated at a location shifted by $\Delta\theta$ to account for the lensing magnification, phase shift, and time delay. The multidimensional posterior distributions themselves are reconstructed from already available posterior samples using a density estimation algorithm, specifically, with \texttt{scipy}'s \texttt{gaussian\_kde}.

Barsode et al.~\cite{barsode2026lensing} made a \textit{B-B} plot for the PO2.0 $\bIII$, and found that it was an underestimate by a factor of $\sim 16$. Extensive code review revealed a few missing factors of 2 in the implementation, which were subsequently fixed. This highlights the potential of \textit{B-B} plots at identifying human errors.

Nevertheless, the corrected version of the code still showed a \textit{B-B} plot biased by $\sim 2$ (Fig.~5 of \cite{barsode2026lensing}). This was puzzling since the approximation made in Eq.~\eqref{eq:GWSL_BLU_joint_likelihood_PO2} is expected to hold whenever the noise across different signals is uncorrelated (i.e., for non-overlapping signals), and when sub-dominant modes of gravitational radiation are weak or absent. Both of these conditions are satisfied by GW signals observable by the current generation of ground-based detectors, implying that the source of this factor of 2 was not theoretical, but numerical. Indeed, further analysis showed that this discrepancy ranged between a factor of 2 on \textit{either side} of equality as we changed the bandwidth or the parameterization of the \texttt{gaussian\_kde}, making it the likely culprit.

In high-dimensional distributions having non-linear correlations and multimodalities-- such as those commonly observed in GW posteriors-- density estimation is prone to biases. Kernel-based methods can be especially limited due to over- or undersmoothing, or ``spillover'' outside the prior boundaries. Recent machine learning techniques, such as normalizing flows, could potentially mitigate these issues by learning a series of invertible transformations between the given samples and a multivariate Gaussian distribution. A sufficiently deep network can fit GW posteriors with remarkable accuracy. Furthermore, once fit, these flows are quite inexpensive to evaluate and resample from.

Here, we consider a normalizing flow implementation called \texttt{denmarf}~\cite{lo2023denmarf}. In Fig.~\ref{fig:test_kde_denmarf}, we show a comparison of density estimation using \texttt{gaussian\_kde} against \texttt{denmarf}. This comparison is made using a randomly generated Gaussian mixture having qualitative features similar to real GW posterior distributions ($\sim 8$ dimensions, non-linear correlations, multimodality, prior boundary effects, etc.), while remaining analytically tractable. We find that \texttt{denmarf} produces a smaller bias and variance in the residuals of the estimated density as compared to \texttt{gaussian\_kde}. We also test different combinations (4,8), (8,16), and (16,32) of the flow's number of hidden units and blocks, and find the residuals to be similar among them.

Since \texttt{denmarf}'s density reconstruction is unbiased and accurate, integrals based on these reconstructions, such as those in Eq.~\eqref{eq:GWSL_BLU_joint_likelihood}, are expected to be accurate as well. Thus, we present a new implementation of PO2.0 based on \texttt{denmarf}. We choose a flow with 8 hidden units and 16 blocks as a middle-ground between computational costs and the flexibility in modeling potentially more complicated distributions than what we tested. The reconstruction is performed in the standard BBH parameter space modified by logit transformations (natively implemented in \texttt{denmarf}) needed to manage prior boundary effects. For a better reconstruction, we explicitly remove the $m_1=m_2$ boundary (where $m_1\geq m_2$ are the component masses of the BBH) by mirroring the posterior distributions across the $m_1=m_2$ line, and account for it by multiplying the estimated density by a factor of 2. Each flow is trained for 1000 epochs, at which point the likelihood converges in all of our tests.

We validate the new PO2.0 using simulated lensed and unlensed GW signals (see Appendix~\ref{sec:PO2_kde_vs_denmarf} for simulation details) and produce a \textit{B-B} plot. As Fig.~\ref{fig:PO2_kde_vs_denmarf_BB_plot} shows, the \texttt{denmarf} based implementation is indeed less biased than \texttt{gaussian\_kde}, with badness values $b=3.2\sigma$ and $7.5\sigma$ respectively. The remaining systematic uncertainty is now \textit{below} the inherent statistical uncertainty due to a finite number of posterior samples (not shown in Fig.~\ref{fig:PO2_kde_vs_denmarf_BB_plot}. See \cite{barsode2025fast}, and also Appendix~\ref{sec:PO2_kde_vs_denmarf}). Additional comparisons between the old and the new method can be found in Appendix~\ref{sec:PO2_kde_vs_denmarf}.

We conclude that the new PO2.0 implementation computes the $\bIII$ with high accuracy, as backed by the \textit{B-B} plot. Furthermore, it is also about one to two orders of magnitude less computationally expensive, marking a significant step towards detecting strongly lensed GWs within realistic computational budgets.

\section{Estimating Backgrounds}
\label{sec:background_estimation}
In the previous section, we demonstrated two examples for benchmarking the accuracy of Bayes factors using backgrounds and foregrounds. For some, these validated Bayes factors may be sufficient at their face value for model selection, while others may demand a comparison with a background distribution for estimating the statistical significance of that result.

In many realistic problems, background generation can be highly computationally expensive. The purpose of this section is to illustrate the power of \textit{B-B} plots at estimating backgrounds at low computational costs. The implicit assumption here is that the Bayes factors are accurate and that all of our modeling assumptions are close to reality so that the \textit{B-B} relationship holds.

This is arguably rather idealistic to assume, and real observations will still demand a rigorous background simulation. However, a simple estimate of the statistical significance can be valuable for an initial assessment of an outlier, and it can be used to justify/reject detailed investigations. The methods outlined here may also be useful for validating search pipelines in regimes not directly accessible due to computational limitations, as well as for forecasting purposes.

We briefly note that while our main focus will be on background estimation, similar techniques can also be applied for estimating foregrounds. This may have some applications in hierarchical search pipelines (eg.~\cite{wright2025lensingflow}) for choosing intermediate thresholds, etc.

\subsection{Background for Strong Lensing of GWs}

\begin{figure*}
\centering
\includegraphics[width=0.8\linewidth]{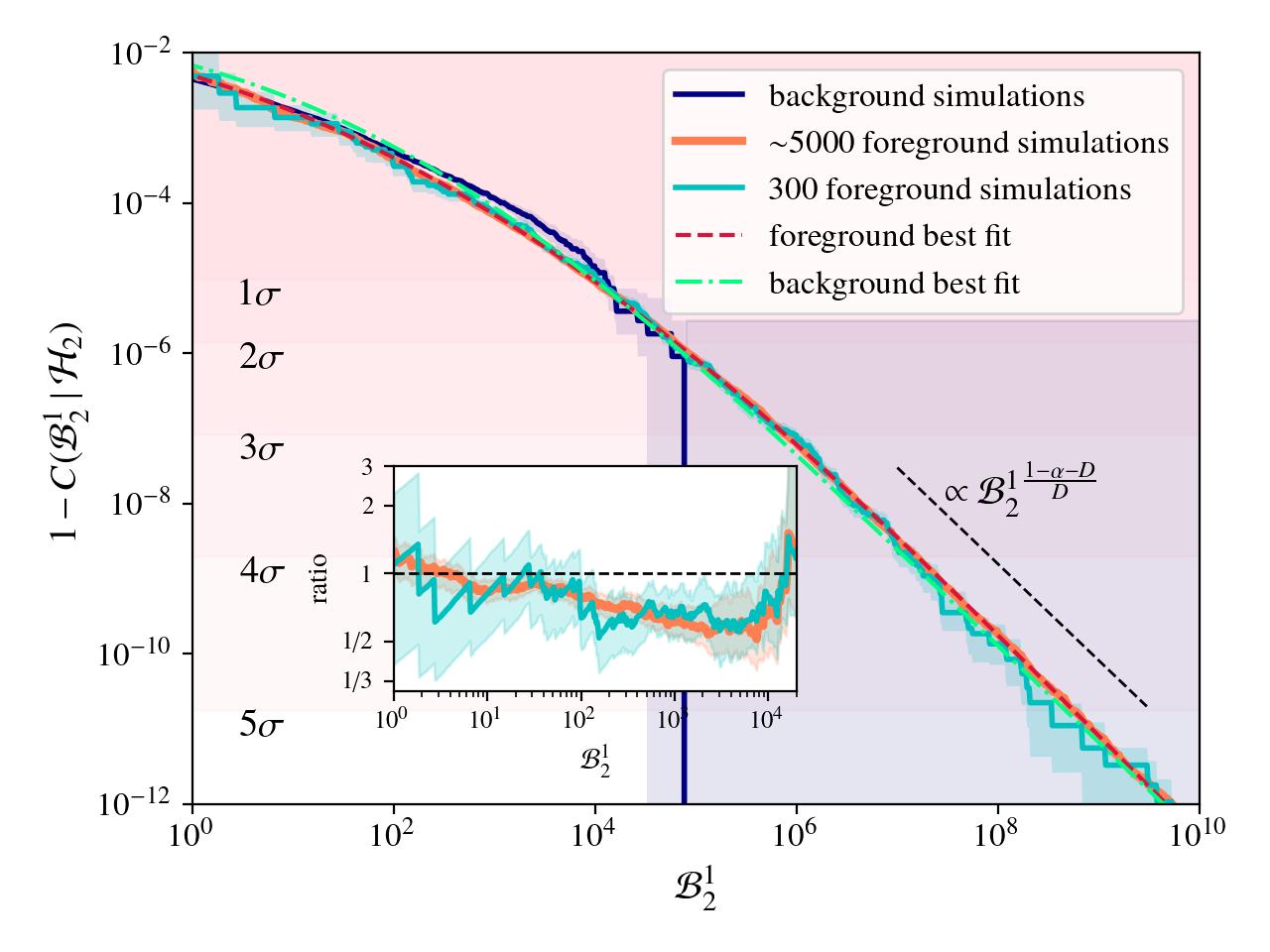}
\caption{The (pairwise) false positive probability (i.e., the survival function of the background) of the GW strong lensing Bayes factor. This is shown for ground truth background simulations, as well as for estimates made using foreground samples weighted according to the \textit{B-B} relationship, and semi-empirical fits to the foreground and background. Ratios of these estimates against the ground truth are shown in the inset, up to the region where the latter is valid. Shaded regions around the histograms indicate 90\% credible regions estimated with bootstrapping, while pink shades indicate statistical significance $\sigma$ thresholds in a catalog of 254 events. Dashed line indicates a power law behavior of the background, with power law index related to the slope $\alpha$ of the SNR distribution, as well as the effective dimensionality (D) of the posterior distributions (see Eq.~\eqref{eq:GWSL_BLU_bg_large_blu}).}
\label{fig:bg_from_fg}
\end{figure*}

Because the search for strongly lensed GWs must be performed over all possible pairs of signals in the catalog, the computational costs rise quadratically with the size of the catalog. The background needs to simulate roughly a few million \textit{times} more pairs to enable establishing statistical significance up to $5\sigma$. For the $\sim 250$ GW events detected so far~\cite{gracedb-o4}, this is $\sim 10^{11}$ pairs, requiring about a year of computation on a large cluster to finish calculating $\bIII$'s using the PO2.0 method (see Sec.~\ref{sec:benchmarking_PO2}).

By the time when the first detection is likely to occur (the fifth observing run (O5) of LIGO-Virgo-KAGRA (LVK)~\cite{barsode2026lensing}), the costs will be even higher, by about three to four orders of magnitude. Thus, with brute force becoming forbiddingly expensive, there is clear motivation to look for alternative ways to estimate the background distribution.

The crux of the problem is that while we need to probe the background at the high-$\bIII$ tail (where lensed events and false alarms would lie), by virtue of it \textit{being} the tail, it is highly inefficient to sample from it using a brute force approach. The \textit{B-B} relationship provides a potential way to improve the sampling efficiency: we can simulate a \textit{foreground} (which samples precisely the range of interesting $\bIII$'s), and convert it to a background through Eq.~\eqref{eq:general_BB}~\cite{barsode2026lensing}.

The conversion can be done by dividing the empirically constructed (e.g., using a histogram) probability distribution $P(\bIII \mid \HI)$ by $\bIII$. However, since samples of the lensed $\bIII$'s are available, the same can be achieved more robustly by reweighting each sample by its inverse, leading to (pairwise) FPP
\begin{equation}
\label{eq:GWSL_bg_from_fg}
1-C(\bIII \mid \HII) \approx \dfrac{1}{N_\ell} \sum\limits_{k=1}^{N_\ell} \dfrac{\Theta(\bIII{}_k \geq \bIII)}{\bIII{}_k}
\end{equation}
where $N_\ell$ is the number of samples $\{\bIII{}_k\}$ available from the foreground $P(\bIII \mid \HI)$ and $\Theta$ is the Heaviside step function. The ratio of estimated FPPs is plotted in the inset in Fig.~\ref{fig:bg_from_fg}, and shows that, at current detector sensitivities, this approach can estimate FPPs accurate to within a factor of 2 using just 300 foreground simulations. This is to be contrasted against the $\sim 10^{11}$ simulations needed in the brute force approach.

\subsubsection{A semi-empirical approach}
Here, we show that it is possible to obtain a deep background semi-empirically, using a smaller background or foreground. The idea is as follows: there exist theoretical or empirically observed correlations between the foreground Bayes factor and certain properties of the signals (e.g., the SNR). It is easy to obtain the distribution of these properties using theoretical arguments or simulations, using which one can marginalize over them to obtain the foreground distribution of the Bayes factor. The \textit{B-B} relationship is then used to transform it to the background distribution. There may be unknown parameters describing such backgrounds, which need to be estimated by fitting with a few simulations.

In the case of strong lensing of GWs, Barsode et al.~\cite{barsode2026lensing} showed that lensed $\bIII$ are correlated with a particular combination of SNRs $\rho_1, \rho_2$ of the two lensed images
\begin{equation}
\label{eq:GWSL_BLU_SNR_correlation}
\log_{10}\bIII = D\log_{10}\rho + \mathcal{N}(\mu, \sigma)
\end{equation}
where $D$ is the effective dimensionality of the posterior distributions of individual signals, $\mathcal{N}(\mu, \sigma)$ describes (log) Gaussian uncertainty with mean $\mu$ (not to be confused with lensing magnification) and standard deviation $\sigma$,\footnote{to be precise, the correlation in Eq.~\eqref{eq:GWSL_BLU_SNR_correlation} as proposed by \cite{barsode2026lensing} excluded time delay information, but here we absorb it into the Gaussian uncertainty} and $\rho = \sqrt{2\rho_A^2\rho_B^2/(\rho_A^2+\rho_B^2)}$.

In the local universe, homogeneity and isotropy would suggest that the distribution of $\rho_A$ and $\rho_B$ should be a power law with slope -4 above a detection threshold $\rho_\mathrm{th} \sim 8$.\footnote{For a uniform density of GW sources, the distance $d$ would be distributed as $\propto d^2$. Since $\rho\propto1/d$, including the Jacobian $\propto \rho^{-2}$, we obtain $P(\rho)\propto\rho^{-4}$.} Strictly speaking, since the sources observable by the LVK are already at cosmological scales, this is not expected to be true. Further, $\rho$ will be distributed differently from $\rho_A$ or $\rho_B$ even if the relative magnification is close to unity and $\rho_A\simeq \rho_B$. However, the deviation is expected to be small, and if desired, the exact distribution can be easily incorporated using astrophysical simulations.

Here, as a simple model for demonstration, we consider $\rho$ to be distributed as a power law with slope $-\alpha$ above $\rho_\mathrm{th}$. Then, it is possible to marginalize Eq.~\eqref{eq:GWSL_BLU_SNR_correlation} over $\rho$ to obtain the foreground distribution in terms of $\alpha, \rho_\mathrm{th}, D, \mu$ and $\sigma$
\begin{multline}
\label{eq:GWSL_BLU_SNR_marginalization}
P(\log_{10}\bIII \mid \HI) = (\alpha-1)~\rho_\mathrm{th}^{\alpha-1} \int\limits_{\rho_\mathrm{th}}^{\infty} \mathrm{d}\rho ~ \rho^{-\alpha}\\
\times \mathcal{N}(\mu + D\log_{10}\rho - \log_{10}\bIII, \sigma).
\end{multline}
This can be integrated analytically, resulting in
\begin{multline}
\label{eq:GWSL_BLU_fg}
P(\log_{10}\bIII \mid \HI) = \dfrac{1}{2}\beta\rho_{th}^{\alpha-1} ~ e^{\beta\mu + \frac{1}{2}\beta^2\sigma^2} ~ \bIII^\frac{1-\alpha}{D}\\
\times \left(1 - \mathtt{erf}\left(\dfrac{\beta \sigma}{\sqrt{2}} + \dfrac{\mu + D\log_{10}{\rho_{th}}}{\sqrt{2}\sigma} - \dfrac{\log_{10} \bIII}{\sqrt{2}\sigma}\right)\right)
\end{multline}
where $\beta = \log{10} \times (\alpha-1) / D$. The background is then given by the \textit{B-B} relationship (Eq.~\eqref{eq:general_BB}, with an extra $1/\bIII$ Jacobian to transform from $P(\log_{10}\bIII)$ to $P(\bIII)$)
\begin{equation}
\label{eq:GWSL_BLU_bg}
P(\bIII \mid \HII) = \dfrac{P(\log_{10}\bIII \mid \HI)}{\bIII{}^2}.
\end{equation}

At large $\bIII$, the $\mathtt{erf}$ term is insignificant, implying that the background, and consequently the FPP in strong GW lensing, is a power law in $\bIII$
\begin{equation}
\label{eq:GWSL_BLU_bg_large_blu}
1 - C(\bIII \mid \HII) \propto \bIII{}^\frac{1-\alpha-D}{D}.
\end{equation}

While these formulae are completely analytical, the parameters $\alpha, \rho_\mathrm{th}, D, \mu, \sigma$ must be anchored to data by fitting Eq.~\eqref{eq:GWSL_BLU_fg} to available foreground, or Eq.~\eqref{eq:GWSL_BLU_bg} to available background. The latter is a significant result: it lets us take a limited background and extrapolate it without running any extra simulations.

For the PO2.0 data from Sec.~\ref{sec:benchmarking_PO2} and Appendix~\ref{sec:PO2_kde_vs_denmarf}, we find the best fit $[\alpha, \rho_\mathrm{th}, D, \mu, \sigma]$ to be $[3.7, 8.2, 8.9, -4.6, 1.5]$ for the foreground, and $[3.5, 8.0, 9.0, -5.0, 1.4]$ for the background. The analytical FPP vs $\bIII$ curves corresponding to these best fit parameters are shown in Fig.~\ref{fig:bg_from_fg} and match within a factor of 2 with the brute force background, and quite closely with the sampling-based background obtained from foreground using Eq.~\eqref{eq:GWSL_bg_from_fg}.

\subsection{The Lensing Interpretation of GW231123}
\label{sec:GW231123}

\begin{figure*}
\centering
\includegraphics[width=\linewidth]{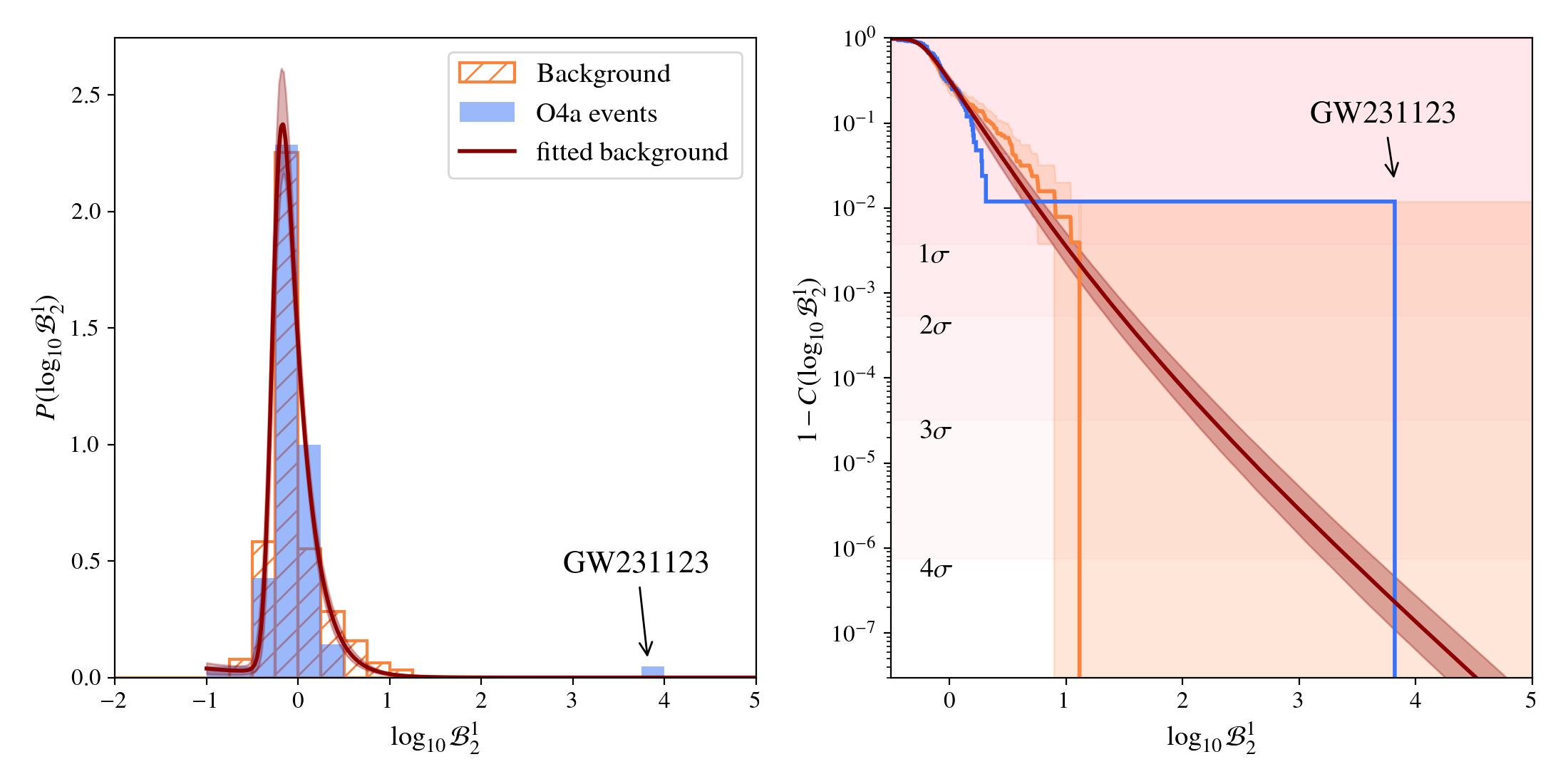}
\caption{Left: histograms of wave-optics lensing Bayes factors for O4a events and background simulations as published by the LVK. Also plotted is our 90\% credible range of possible semi-empirical backgrounds in dark red. Right: The (eventwise) false positive probability (i.e., the survival function of the background distribution), along with a survival function of the observed events, and our semi-empirical extrapolation to the background. Orange shaded regions show 90\% credible regions estimated with bootstrapping, while pink shades indicate statistical significance $\sigma$ thresholds in a catalog of 84 events.}
\label{fig:GW231123_bg}
\end{figure*}

GW231123 was identified by the LVK~\cite{abac2025gwtclens} as an outlier with a Bayes factor $\sim10^4$ in favor of it being a signal with a waveform modified due to lensing in the wave-optics regime. If GW231123 is indeed lensed, establishing $5\sigma$ statistical significance would require $\gtrsim 10^8$ background simulations\footnote{including the trials factor corresponding to 84 events in O4a}. However, with only 254 simulations in their computationally limited background, the LVK could only put a lower bound of $\sim 1\sigma$. While non-lensing explanations such as overlapping signals~\cite{hu2025gw231123} or a merger of primordial black holes~\cite{de2025gw231123, yuan2025gw231123} have been proposed, a number of lensing studies have also followed~\cite{goyal2025across, chan2025discovering, chakraborty2025first, shan2025gw231123, wang2026gw231123}. Chan et al.~\cite{chan2025discovering}, in particular, used simulation-based inference to generate a deep, brute-force background to establish an upper bound of $\sim 4\sigma$.

Here, we will obtain the same result, but instead of running a large number of simulations, we will extrapolate LVK's limited astrophysical background using a semi-empirical approach based on the \textit{B-B} relationship similar to that shown in the previous subsection. Since this implicitly assumes stationary Gaussian noise, our answer will only be a rough, order-of-magnitude estimate.

First, we need a prescription for the foreground. We use the ``fitting factor'' approximation given by~\cite{cornish2011gravitational, vallisneri2012testing, pozzo2014testing}. The expectation is that, in the high-SNR limit, the logarithm of the Bayes factor increases quadratically with the SNR $\rho$. Its coefficient is related to a ``fitting factor'' $FF \in [0,1]$ that quantifies how well an unlensed waveform can match a lensed waveform. Including another term proportional to $\log{FF}$~\cite{pozzo2014testing} and a $\sim \rho\sqrt{2(1-FF)}$ Gaussian fluctuation~\cite{vallisneri2012testing}, this becomes
\begin{multline}
\label{eq:fitting_factor_blu}
\log \bIII(\rho, FF) \simeq \dfrac{1}{2}(1 - FF^2)\rho^2 + (D - d)\log{FF}\\
+ d\log\rho - \mathscr{O} + \mathcal{N}\left(0, \rho\sqrt{2(1-FF)}\right)
\end{multline}
where $D$ is the number of parameters that we can effectively measure from unlensed signals, $d$ is the number of additional parameters required for describing the lensed signal, and $\mathcal{N}(\mu,\sigma)$ denotes Gaussian uncertainty with mean $\mu$ and standard deviation $\sigma$. $\mathscr{O}$ is the Bayesian Occam factor that quantifies the reduction in the prior volume of the additional parameters due to the information in the data. In general, the measurement uncertainties in each parameter reduce roughly in proportion to $\rho$. We have factorized that $d\log\rho$ dependence out of the Occam factor, making $\mathscr{O}$ roughly constant for a given problem.

Since we are only interested in a rough estimate, we simplify the above even further by effectively marginalizing over the unknown distribution of $FF$. We assume that the $\log_{10}$ Bayes factor is distributed around a mean function $f\rho^2+d\log\rho-\mathscr{O}$, with uncertainties due to those in $FF$ absorbed by an effective standard deviation of the Gaussian
\begin{multline}
\label{eq:fitting_factor_blu_simple}
\log_{10} \bIII(\rho) \simeq f\rho^2 + d\log_{10}\rho - \mathscr{O}\\
+ \mathcal{N}\left(0, s_0+s_1\rho+s_2\rho^2\right).
\end{multline}

Following similar steps as in the previous subsection, we assume a power law distribution for $\rho$ with slope $-\alpha$ above a threshold $\rho_\mathrm{th}$, and numerically marginalize over $\rho$. This results in a semi-empirical foreground distribution, which can then be converted to a background using the \textit{B-B} relationship (Eq.~\eqref{eq:general_BB} or ~\eqref{eq:GWSL_BLU_bg}).

\begin{table}[h]
\begin{tabular}{lllll}
                   & distribution & minimum   & maximum &  \\ \cline{1-4}
$f$                & LogUniform   & $10^{-8}$ & 0.5     &  \\
$\alpha$           & Uniform      & 3         & 5       &  \\
$\rho_\mathrm{th}$ & Uniform      & 7         & 9       &  \\
$\mathscr{O}$      & Uniform      & -10       & 10      &  \\
$d$                & Uniform      & 0         & 3       &  \\
$s_0$              & LogUniform   & $10^{-6}$ & 1       &  \\
$s_1$              & LogUniform   & $10^{-6}$ & 1       &  \\
$s_2$              & LogUniform   & $10^{-6}$ & 1       &
\end{tabular}
\caption{Priors used in fitting unknown parameters of our semi-empirical model to LVK's limited background for wave-optics lensing distortion.}
\label{tab:GWTC4_bg_fit_priors}
\end{table}

We constrain the free parameters $f, \alpha, \rho_\mathrm{th}, d, \mathscr{O}, s_0, s_1, s_2$ by fitting our semi-empirical model to LVK's background sample of 254 simulations. For each point in this parameter space, we obtain $P(\log_{10} \bIII \mid \HI)$ by numerically marginalizing Eq.~\eqref{eq:fitting_factor_blu_simple} over $\rho$, and convert it to $P(\log_{10} \bIII \mid \HII)$ by dividing by $\bIII$. The likelihood is then given by the product $\prod_i P(\log_{10} \bIII_i \mid \HII)$, where $\{\bIII_i\}$ are the 254 background Bayes factors provided by the LVK. Using the priors listed in Tab.~\ref{tab:GWTC4_bg_fit_priors}, we sample the posterior with the \textsc{Nautilus}~\cite{nautilus} nested sampler using 1024 live points, run until the fractional evidence in the live set is $\leq 1\%$, or the effective number of samples is $\geq 10^4$.

Once posterior samples are obtained, we can compute the corresponding $P(\log_{10} \bIII \mid \HII)$, and numerically integrate them to obtain the reversed CDFs. The resulting range of possible backgrounds is shown in Fig.~\ref{fig:GW231123_bg}. Nominally, we find that GW231123 is a 4.1-4.5$\sigma$ significant candidate for wave-optics lensing.

To check the robustness of these fits, we experimented with prior ranges different from those listed in Tab.~\ref{tab:GWTC4_bg_fit_priors}, as well as with fits where one or two of $s_0,s_1,s_2$ were fixed at 0. Some of these choices extended possible backgrounds down to slightly lower FPPs, though the estimated statistical significance remains $\gtrsim 4.1\sigma$. Further validation of this methodology is presented in Appendix~\ref{sec:dingo_lensing}.

A confident interpretation-- lensing or otherwise-- is hampered by the presence of severe waveform systematics~\cite{abac2025gw231123, bini2026impact, chan2025discovering} and probable noise transients within the event's strain~\cite{chatterjee2025machine, ray2025gw231123, bini2026impact}, both of which are outside the scope of a \textit{B-B} plot based background estimate. The \textit{B-B} plot assumes stationary Gaussian noise (same as the likelihood used in the computation of evidences), while background injections in real noise are prone to non-stationarities as well as non-Gaussianities. These could result in the \textit{real} background having heavier tails than our Gaussian expectation~\cite{sasli2023heavy}. This is perhaps visible in our fits: while they remain consistently within errorbars of the real tail, they appear to lie on the lower side.

Thus, our estimate of the statistical significance should be treated as an upper bound of $4.1\sigma$. This is consistent with that estimated by \cite{chan2025discovering} using a detailed study, though we remark that their background was not distributed according to astrophysical priors, and it did not include the catalog trials factor of $\sim 84$.

For more impactful estimates of the statistical significance, some improvements would be desirable in our semi-empirical background model (Eq.~\eqref{eq:fitting_factor_blu_simple}). For instance, we can use simulated distributions of fitting factors and SNRs for marginalization, or the Gaussian noise term could be substituted with a heavier-tailed distribution to model the low-FPP region more accurately. We would also need to include selection effects due to the initial matched filtering step that is usually performed without taking waveform distortion into account~\cite{chan2025detectability}. Importantly, these estimates would need calibration against at least a limited but real background. We leave these as future work.

\section{Conclusion}
\label{sec:conclusion}
Bayes factors are foundational to many scientific discoveries in the modern era. Yet, due to computational complexity, the practical implementation of Bayes factor calculation may be only approximate. There is also a possibility of human error, resulting in potentially biased estimates.

In this paper, we have described the Bayes factor-Bayes factor or \textit{B-B} plot to assess the accuracy of such implementations without requiring access to the ground truth. Guided by \textit{B-B} plots, we improved the PO2.0 method for searching strongly lensed GWs, with the new implementation being both accurate and more computationally efficient. It may be instrumental in realizing the detection of lensed multiple images anticipated in the upcoming years, especially in the face of large catalog sizes.

Using the \textit{B-B} plot, we also showed that it is possible to detect gravitational waveform distortion using fast Fisher covariance matrix-based Bayes factors. Here, we used a toy model for demonstration, but we are currently working on applying these methods in realistic searches for wave-optics and saddle-point lensing of GWs, tests of GR, orbital eccentricity, and environmental effects.

We note that the \textit{B-B} plots offer a necessary consistency test, but they are not sufficient. Arbitrary likelihood ratios may satisfy them, though we believe this should be recognizable in the practical implementation. Researchers may judge whether their computation of the Bayes factor indeed involves a ratio of evidences under the two hypotheses, or is merely a likelihood ratio of arbitrary statistics derived from the data. Further work may focus on removing this caveat with help from other simulation-based validation methods proposed in the literature~\cite{modrak2023simulation, modrak2025simulation}.

Beyond diagnostics, the \textit{B-B} relationship can also be used to estimate frequentist style backgrounds using the Bayes factor as the optimal statistic. We derived an analytical expression for the background of GW strong lensing, and showed that it can estimate FPPs within a factor of 2 of the ground truth. Being analytical, these backgrounds can be easily extrapolated to FPPs that are unreachable by brute force simulations. Using a similar approach but in the wave-optics regime of lensing, we also derived a semi-numerical extrapolation for the background in GWTC4, and used it to estimate a rough bound on the statistical significance of the lensing hypothesis of GW231123 at $\lesssim 4.1\sigma$, assuming Gaussian noise.

These semi-empirical background-foreground models open a rather intriguing possibility\footnote{partially credited to Otto A. Hannuksela}: we may fit the distribution of real events' Bayes factors with a mixture of background and foreground models. The mixture coefficient can be estimated from the real data itself, resulting in bounds on the presence and statistical significance of outliers without requiring any simulations. Future work would be to test how reliable such estimates are in comparison to rigorous studies.

While the statistical significance of an observed candidate will ultimately be decided by rigorous background simulations regardless of the computational costs, the rough bound estimated with our approach is still valuable for an initial assessment of outliers similar to GW231123. Furthermore, these techniques will certainly be useful in the development of search pipelines and in making forecasts.

The applications discussed here were limited to GW astronomy, however \textit{B-B} plots are quite general and could be useful in any field that relies on model selection with Bayes factors. While their effectiveness depends on the accuracy of the underlying statistical assumptions, they offer a simple and computationally efficient tool for both validating Bayes factor calculations and estimating statistical significance. We anticipate that these methods will find broader use in large-scale inference problems where exact calculations are impractical.

\begin{acknowledgments}
I thank Juno Chan for graciously providing data from their work on \textsc{dingo-lensing}~\cite{chan2025discovering, chan2026identification}. I also thank Mick Wright, Ajith Parameswaran, and Anjali Bhatter for a careful review of this manuscript. I acknowledge support from Koustav N. Maity, Siva Athreya, Otto A. Hannuksela, Rico Lo, V Sree Suswara, and Dootika Vats. I also thank members of the astrophysical relativity group at ICTS and the LVK Lensing group for constructive feedback on this work. My research is supported by the Department of Atomic Energy, Government of India, under Project Nos. RTI4019 and RTI4013. The numerical calculations reported in the paper were performed on the Alice computing cluster at ICTS-TIFR.

This work makes use of the \texttt{cogwheel} \cite{roulet2022removing, islam2022factorized}, \texttt{scipy} \cite{2020SciPy-NMeth}, and \texttt{nautilus} \cite{nautilus} software packages. This material is based upon work supported by NSF's LIGO Laboratory, which is a major facility fully funded by the National Science Foundation.
\end{acknowledgments}

\section*{Data Availability}
The codes used in this study are available at \texttt{https://github.com/anbarsode/bb\_plot}. The data are available from the author upon request.

\appendix

\section{GW strong lensing search using PO2.0}
\label{sec:PO2_kde_vs_denmarf}

\begin{figure*}
\centering
\includegraphics[width=\linewidth]{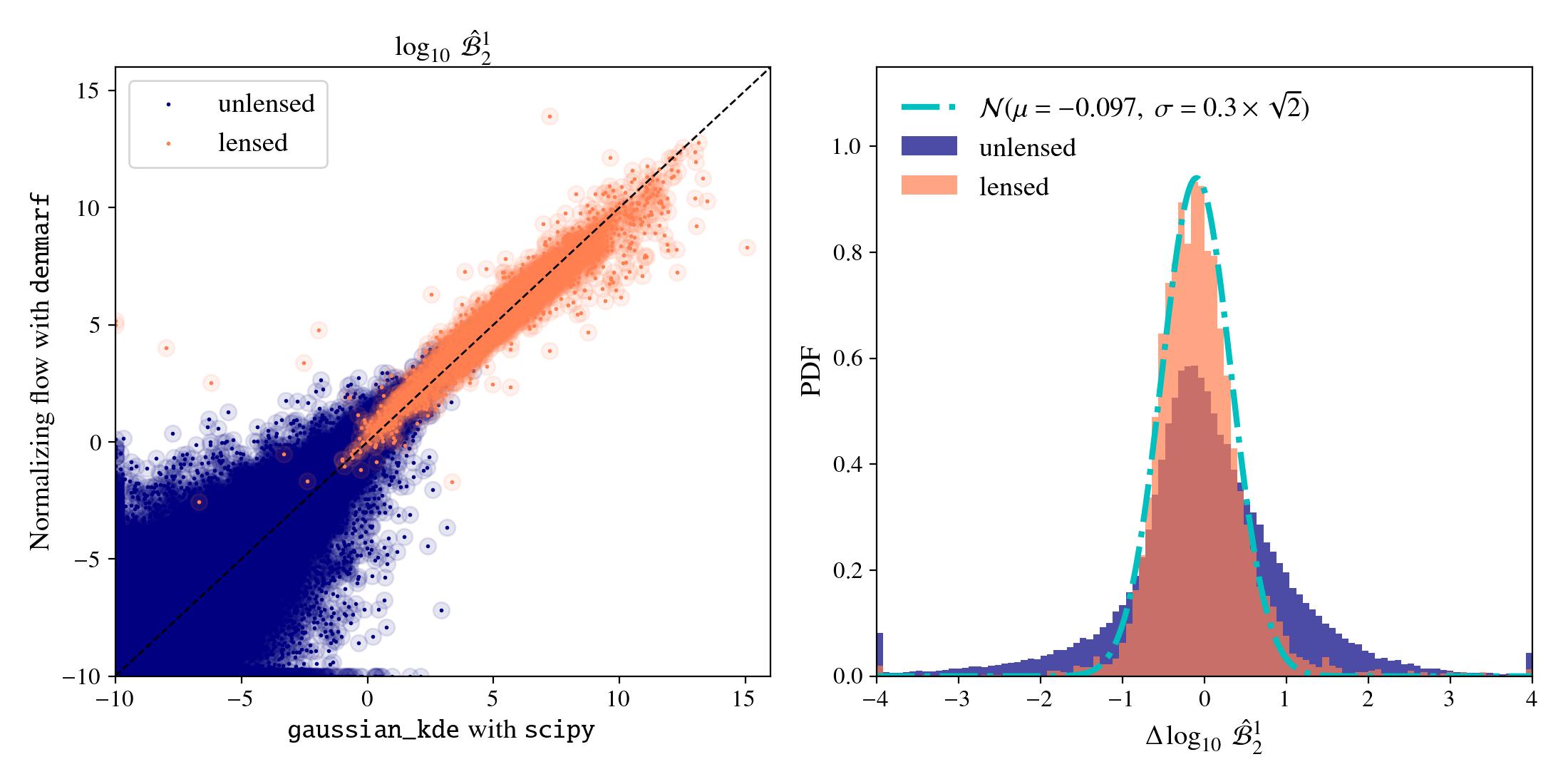}
\caption{Left: the new PO2.0 Bayes factors plotted against the old, both being computed on the same dataset. Faint disks denote 1 standard deviation of statistical uncertainty. Right: histogram of the difference between the new and old implementation. Also plotted is a Gaussian distribution with mean given by the median of the differences between lensed Bayes factors, and a standard deviation estimated using a statistical sampling uncertainty of 0.3 in each measurement~\cite{barsode2025fast}.}
\label{fig:PO2_kde_vs_denmarf}
\end{figure*}

\begin{figure}
\centering
\includegraphics[width=\linewidth]{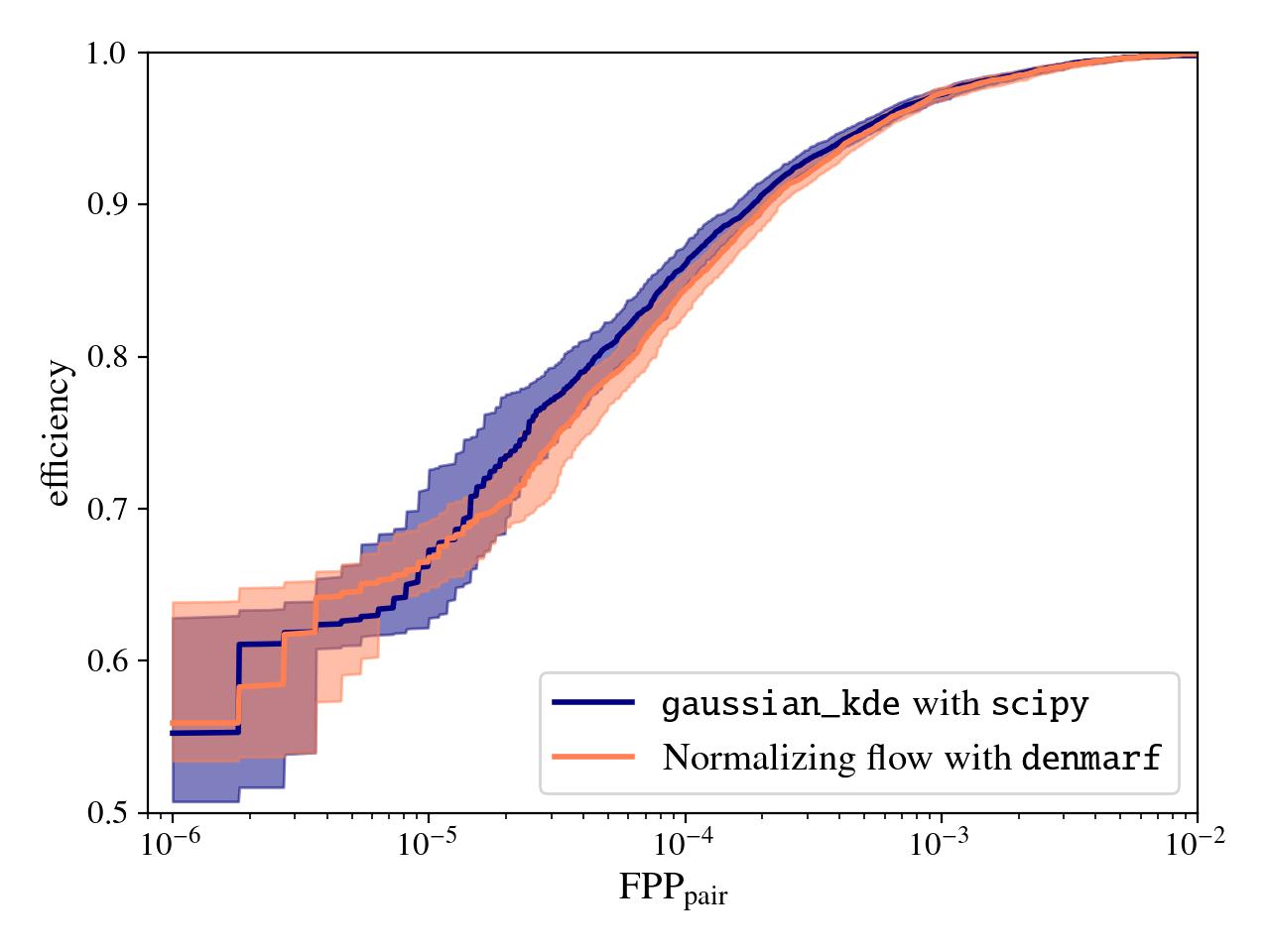}
\caption{The efficiency (i.e., the true positive probability) of finding strongly lensed GWs as a function of the (pairwise) FPP for the old and new implementations of PO2.0. Shaded regions show 90\% credible regions estimated using bootstrapping.}
\label{fig:PO2_kde_vs_denmarf_ROC}
\end{figure}

Here, we outline the methodology used for generating simulated unlensed and strongly lensed GW signals on which we test our implementation of PO2.0. We also provide additional comparison plots between the old and the new implementation.

Our astrophysical simulation is identical to that from \cite{barsode2025fast}, though we use the latest available data regarding the GW populations. We generate an intrinsic population of GW sources according to the population inferred from observed BBHs at the end of the first part of the fourth observing run of the LVK~\cite{abac2025gwtcpop}, with a merger rate following the star formation rate~\cite{madau2014cosmic}. Depending on the optical depth of strong lensing to the GW source's redshift, they are assigned galaxy lenses distributed according to the SDSS catalog~\cite{collett2015population} with a Singular Isothermal Ellipsoid~\cite{kormann1994isothermal, fukugita1991gravitational} profile. Their detectability is assessed by thresholding the optimal SNR at 8, where the SNR is computed at the projected sensitivity corresponding to the fourth observing run (O4) of the LVK~\cite{HLVK-psd-O3O4O5}, assuming a 2.5 year observing duration and ignoring detector down times, using the \textsc{IMRPhenomXPHM}~\cite{pratten2021computationally} waveform model.

With the detectable populations generated as above, 1500 unlensed and 5000 lensed pairs of waveforms are injected in colored Gaussian noise corresponding to the same O4 sensitivity. The effect of Morse phase shift~\cite{dai2017waveforms} is included by multiplying the waveforms of saddle point lensed images by $e^{\iota \pi/2}$. Parameter estimation on these mock strains is performed with the \textsc{cogwheel}~\cite{islam2022factorized, roulet2022removing} package, generating the common dataset on which both the implementations of PO2.0 are tested.

Figure~\ref{fig:PO2_kde_vs_denmarf} shows a comparison of the Bayes factors computed by the two approaches for the same pairs of signals. The left-hand plot shows that the old and the new $\bIII$'s are tightly correlated for lensed pairs. The right-hand panel shows that the residuals follow the expected sampling error budget of $\sim0.3$ in $\log_{10}\bIII$ (with a factor of $\sqrt{2}$ to account for the error in individual calculation), though with a small shift and skew, signifying a non-trivial change brought in by the new method. For unlensed pairs, there is a relatively large scatter between the $\bIII$'s computed with the two methods, and it does not follow the estimated error budget. However, since lensing search focuses entirely on the high $\bIII$ regime, this does not concern us.

The accuracy of unlensed pairs' Bayes factors \textit{is} important for false alarms, i.e., unlensed pairs that accidentally result in high $\bIII$'s. This is best assessed with Receiver Operating Characteristics (ROC) plots, showing the efficiency (i.e., the true positive probability) against the (pairwise) FPP. Figure~\ref{fig:PO2_kde_vs_denmarf_ROC} shows that the old and new methods agree to within the expected sampling error budget.

\section{Validating \textsc{dingo-lensing} and Background Estimation}
\label{sec:dingo_lensing}
\begin{figure}
\centering
\includegraphics[width=\linewidth]{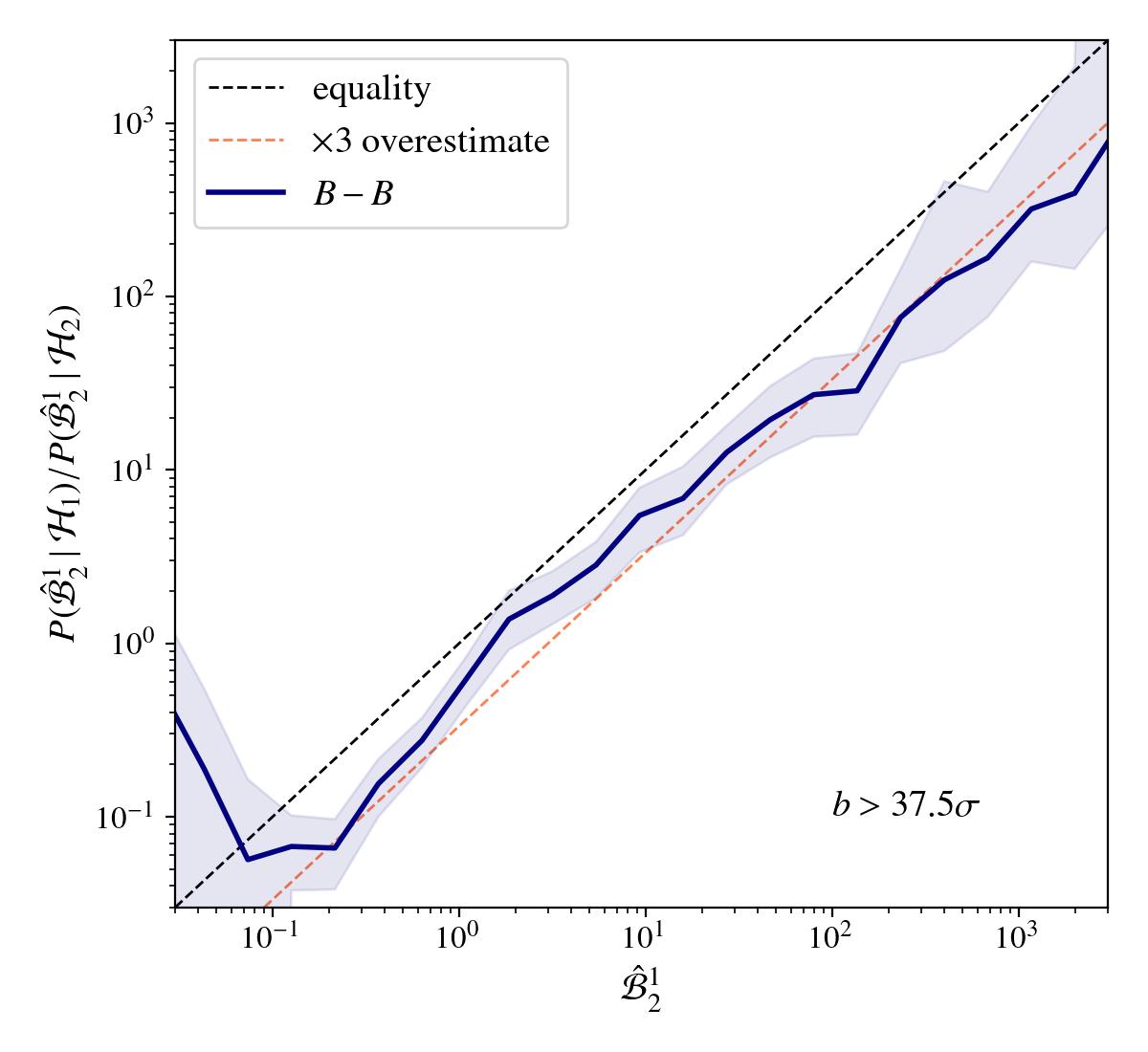}
\caption{The \textit{B-B} plot made for \textsc{dingo-lensing} Bayes factors made using data from \cite{chan2025discovering} for the \textsc{NRSur7dq4} waveform model. Shaded regions indicate 90\% credible regions estimated assuming multinomial histogram errors.}
\label{fig:Juno_GW231123_BB_plot}
\end{figure}

\begin{figure*}
\centering
\includegraphics[width=\linewidth]{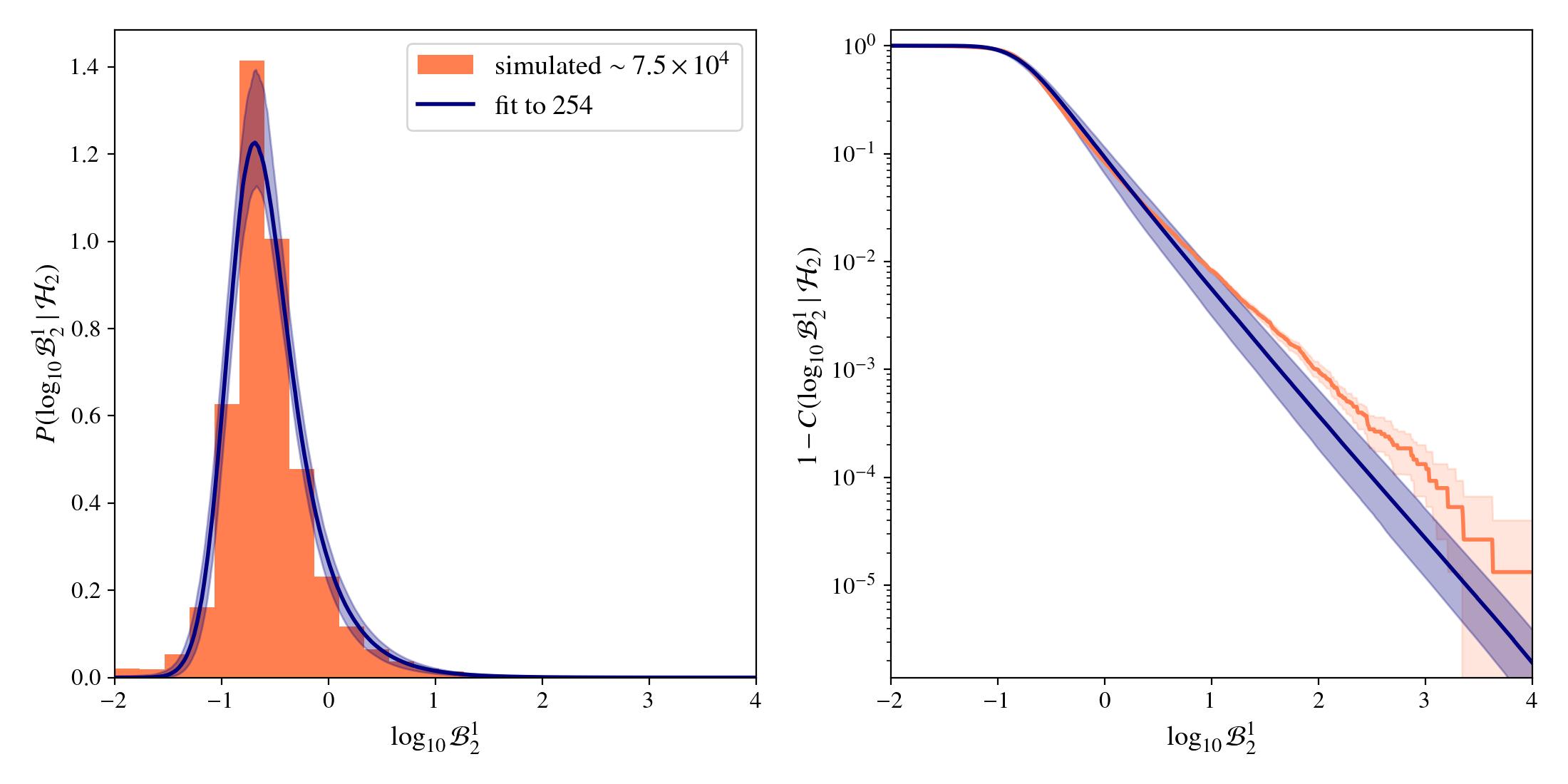}
\caption{Histogram (left) and reversed cumulative distribution (right) for \textsc{dingo-lensing} Bayes factors from \cite{chan2025discovering} for the \textsc{NRSur7dq4} waveform model. Also plotted is a semi-empirical fit based on Eq.~\eqref{eq:fitting_factor_blu_simple} to a smaller sample of 254 to show that the extrapolation bounds the cumulative distribution from below. Shaded regions indicate 90\% credible regions.}
\label{fig:Juno_GW231123_fit}
\end{figure*}

Chan et al.~\cite{chan2026identification, chan2025discovering} developed a neural posterior estimation framework called \textsc{dingo-lensing} for studying lensing of gravitational waves. This code drastically reduces the computational cost of computing lensing Bayes factors, making it possible to generate foregrounds and backgrounds within realistic computational budgets.

In particular, Chan et al. produced a background and foreground for GW231123-like signals using the \textsc{NRSur7dq4} waveform model, shown in Fig.~4 of \cite{chan2025discovering}. Using the original data provided by them, we show a \textit{B-B} plot in Fig.~\ref{fig:Juno_GW231123_BB_plot}, finding a systematic deviation from the \textit{B-B} relation by a factor of $\sim$3 (badness more than floating point resolution, see Eq.~\eqref{eq:badness}). This is larger than the statistical fluctuation of $\sim 0.7-1.4$ they find when compared against nested sampling using \textsc{bilby}. The higher systematic deviation is likely caused by a mismatch in the priors used for drawing the injection parameters and those used for evaluating the Bayes factors (the latter being broader than the former), which breaks one of the requirements of making the \textit{B-B} plot. We emphasize that this \textit{does not alter their conclusions in \cite{chan2025discovering}}, since they were made using frequentist estimates that are robust against such mismatches.

We next test whether our semi-empirical formulation (in particular, Eq.~\eqref{eq:fitting_factor_blu_simple}) for the background of waveform distortion effects matches Chan et al's distributions, or at least bounds the FPPs from below. Since the \textit{B-B} plot is sensitive to waveform systematics, we restrict this test to \textsc{NRSur7dq4}, which is the most appropriate among the available waveform models in the parameter range surrounding GW231123~\cite{abac2025gw231123}. Note that the above-mentioned bias would not affect this fit; it simply gets absorbed in the free parameter $\mathscr{O}$.

We draw 254 random samples from their background Bayes factors, fit them with the semi-empirical procedure described in Sec.~\ref{sec:GW231123}\footnote{with priors same as those in Tab.~\ref{tab:GWTC4_bg_fit_priors} except for a more relaxed prior for $\alpha\in[1,5]$}, and plot its extrapolation alongside the reversed cumulative histogram obtained from the full dataset of $\sim 7.5\times 10^4$. As Fig.~\ref{fig:Juno_GW231123_fit} shows, the semi-empirical fit provides a consistent lower bound, validating the results in Sec.~\ref{sec:GW231123}.

\bibliography{bibliography}% Produces the bibliography via BibTeX.

\end{document}